\documentclass[11pt,a4paper,dvips]{article}            

\usepackage{a4wide}
\usepackage[english]{babel}
\usepackage{epsfig}

\newcommand{\be}{\begin{eqnarray}}
\newcommand{\ee}{\end{eqnarray}}
\newcommand{\beqn}{\begin{eqnarray}}
\newcommand{\eeqn}{\end{eqnarray}}
\newcommand{\bes}{\begin{eqnarray*}}
\newcommand{\ees}{\end{eqnarray*}}
\newcommand{\beqns}{\begin{eqnarray*}}
\newcommand{\eeqns}{\end{eqnarray*}}

\newcommand{\rmd}{\mbox{d}}
\newcommand{\rme}{\mbox{e}}
\newcommand{\rmi}{\mbox{i}}

\newcommand{\srmi}{\mbox{\scriptsize i}}

\newcommand{\pdd}[2]{{\partial{#1}\over\partial{#2}}}

\newcommand{\tr}{\mathop{\mbox{Tr}}}

\newcommand{\erf}{\mathop{\mbox{erf}}}

\newcommand{\sbfit}[1]{\mbox{\scriptsize
\protect\boldmath $#1$}}
\newcommand{\bfit}[1]{\mbox{\protect\boldmath $#1$}}

\newcommand{\ns}[1]{\mbox{\normalsize $#1$}}
\newcommand{\wt}{\widetilde}

\begin{document}

\title{Numerical path integration with Coulomb potential}

\author{Jan Myrheim\\
Department of Physics, NTNU, N--7491 Trondheim, Norway\\
and\\
LPTMS, Bat.\ 100, Universit{\'e} Paris Sud, F--91405 Orsay, France}

\maketitle

\begin{abstract}
A simple and efficient method for quantum Monte Carlo simulation is presented,
based on discretization of the action in the path integral, and a Gaussian
averaging of the potential, which works well e.g.\ with the Coulomb potential.
Non-integrable hard core potentials can not be averaged in the same way.
\end{abstract}

\section{Introduction}

The quantum Monte Carlo method is well established as an efficient
calculational tool for many-body problems.  See for example the review
articles \cite{vdLinden92,Ceperley95,Foulkesetal01}.  It is well
suited for bosonic systems without magnetic field, where the path
integral has only positive contributions.  But also femionic systems
can be treated, in spite of the troublesome ``sign problem'', just one
application is the computation of the high temperature phase diagram
of hydrogen~\cite{MilitzerCeperley01,Filinovetal01}.  One limitation
is that it is a statistical method, so that every factor of 10 in
precision costs a factor of 100 in computing time. But it does not hit
the ``exponential wall'', because it only needs to represent particle
positions, or particle paths, and so the number of parameters
increases linearly with the number of particles.

By contrast, the number of particles that can be handled by methods
based on computing realistic manybody wave functions is limited by
the exponential increase in the number of parameters needed for
describing such wave functions. The density functional method
\cite{Kohn99} is less severely limited, since it uses one particle
wave functions, but on the other hand it has to rely on clever
approximation techniques.

The purpose of the work presented here was to look for a simple and
efficient way of handling the Coulomb potential in quantum Monte Carlo
simulation.  Formulae for the exact propagator of the two-particle
Coulomb problem are known, and can even be derived by path integral
methods~\cite{DuruKleinert82,HoInomata82}, but it is not clear whether
they are useful for the simulation of many-body systems.  In
ref.~\cite{MilitzerCeperley01} fitted formulae for two-particle
propagators were used.  This may be a good enough method, but it may
nevertheless be of interest to look for more direct approaches.

The Fourier representation of paths in path integrals was introduced
by Feynman, together with the idea of approximately integrating over
infinitely many Fourier components by averaging the
potential~\cite{FeynmanHibbs}.  The method was further developed by
Doll, Coalson, and Freeman~\cite{Coalsonetal85,Coalsonetal86} under
the name of partial averaging.  In their work and in subsequent work,
see e.g.~\cite{Chakravarty93,KoledeRaedt01}, the object focused upon
has been the propagator, involving paths from one point to another,
more than the partition function, involving closed paths.

The basic idea is to let the path integral include only paths
represented by finite Fourier series with a fixed number of terms, and
to think of each such path as representing all the infinite Fourier
series to which it can be extended.  The result is that, as long as
correlations along the path are neglected, each point on one truncated
path represents a Gaussian distribution of points, with a standard
deviation decreasing from the middle of the path and vanishing at the
end points.  In the case of the Coulomb potential, the Gaussian
averaging has the important effect of removing the singularity at zero
distance.  It is then a complication that the averaging varies along
the path, disappearing towards the end
points~\cite{KoledeRaedt01}.

The modification proposed here is to average with a standard deviation
which is constant along the path.  This seems a natural approach when
the partition function is computed directly, and not via the
propagator.  In more detail, the method proposed amounts to a
discretization of the action integral, with an averaged potential, and
the computation of the kinetic energy part of the action by means of a
finite Fourier transform.  As discussed below, within this method it
is easy to add to any potential an auxiliary confining harmonic
oscillator potential, which makes the partition function
mathematically well defined.

A different topic which is not addressed here is the optimization of
the Monte Carlo sampling procedure.  See in this connection the
comment in an appendix of ref.~\cite{KoledeRaedt01}.  A finite Fourier
transform is a central part of the present method.  In the method as
formulated here it is assumed that the number of time steps is odd,
hence the standard fast Fourier transform with $2^n$ points,
$n=1,2,\ldots$, can not be used.  One solution is to use $3^n$ points
and the fast Fourier transform to base three~\cite{Pressetal}.
However, it would also be straightforward to modify the method so as
to use an even number of time steps.

\section{The imaginary time path integral}

Equilibrium properties of a physical system at a finite temperature
${\cal T}$ can be computed from the partition function
\be
\label{eq:partf}
Z(\beta)=\tr\rme^{-\beta H}\;,
\ee
where $H$ is the Hamiltonian, and $\beta=1/(k_B{\cal T})$.
One may regard $\beta\hbar$ formally as an imaginary
time interval.

To be specific, we consider most of the time one particle of mass $m$ in three
dimensions. The Hamiltonian is $H=T+V$, with $T=\bfit{p}^2/(2m)$ the kinetic
energy and $V=V(\bfit{r})$ the potential energy. The partition function has
the following path integral representation,
\be
\label{eq:ZnonfreeinfiniteFourier}
Z=
\wt{C}_0^{\;3}\int\rmd^3\!\bfit{a}_0
\prod_{n=1}^{\infty}
\left(\wt{C}_n^{\;6}\int\rmd^6\!\bfit{a}_n\;
\right)
\exp\!\left(-{S\over\hbar}\right).
\ee
We define the constants
\be
C_0={1\over\hbar}\,\sqrt{m\over 2\beta}
\;,\qquad
C_n={2n\pi\over\hbar}\,\sqrt{m\over\beta}
\quad
\mbox{for}\quad n=1,2,\ldots,
\ee
and $\wt{C}_n=C_n/\sqrt{\pi}$ for $n=0,1,2,\ldots$. $S$ is the imaginary time
action,
\be
\label{eq:integrandpathint}
{S\over\hbar}
={1\over\hbar}\int_0^{\beta\hbar}\rmd\tau\left(
{1\over 2}\,m\,(\dot{\bfit{r}}(\tau))^2
+V(\bfit{r}(\tau))\right)
=\sum_{n=1}^{\infty}C_n^{\;2}\,|\bfit{a}_n|^2
+{1\over\hbar}\int_0^{\beta\hbar}\rmd\tau\;V(\bfit{r}(\tau))\;.
\ee
The path $\bfit{r}=\bfit{r}(\tau)$ is periodic in the imaginary time $\tau$,
with period $\beta\hbar$, and is given by the infinite Fourier series
\be
\label{eq:infFourierexp}
\bfit{r}(\tau)
=\sum_{n=-\infty}^{\infty}
\bfit{a}_n\,\rme^{\,\srmi\,{2n\pi\tau\over\beta\hbar}}\;.
\ee
The Fourier components $\bfit{a}_n$ are complex and satisfy the relations
$\bfit{a}_{-n}=\bfit{a}_n^{\ast}$, so that $\bfit{r}(\tau)$ is real. In
particular, $\bfit{a}_0$ is real. The time derivative $\dot{\bfit{r}}$ is with
respect to the imaginary time $\tau$.

This Fourier expansion is the natural one in the computation of the
partition function, which involves periodic paths. A slightly
different expansion is needed in the computation of propagators, see
e.g.\
\cite{FeynmanHibbs,Coalsonetal85,Coalsonetal86,Chakravarty93,KoledeRaedt01}.

It may be useful to sketch the derivation of this Fourier path integral.
We start from the approximation
\be
Z\approx
\tr\prod_{j=1}^J\left(
\rme^{-{\beta\over J}\,T}\,
\rme^{-{\beta\over J}\,V}\right).
\ee
We insert $J$ times the identity operator
$I=\int\rmd^3\bfit{r}\;|\bfit{r}\rangle\langle\bfit{r}|$
where $|\bfit{r}\rangle$ is the position eigenstate,
and introduce the free particle propagator
\be
\langle\bfit{r}'|\rme^{-\beta T}|\bfit{r}\rangle
=\wt{C}_0^{\;3}\,\rme^{-C_0^{\;2}\,|\sbfit{r}-\sbfit{r}'|^2}\;,
\ee
to obtain the approximation
\be
\label{eq:Zprimitive}
Z\approx
\left(\sqrt{J}\,\wt{C}_0\right)^{3J}
\int\rmd^3\bfit{r}_1\,\rmd^3\bfit{r}_2\,\cdots\,\rmd^3\bfit{r}_J\;
\exp\!\left(-{S_P\over\hbar}\right).
\ee
Here $S_P$ is the ``primitive'' discretized action defined by
\be
{S_P\over\hbar}=
JC_0^{\;2}\sum_{j=1}^J|\sbfit{r}_j-\sbfit{r}_{j+1}|^2
+{\beta\over J}\sum_{j=1}^J V(\sbfit{r}_j)\;.
\ee
We define $\bfit{r}_{J+1}=\bfit{r}_1$. This approximate expression for $Z$ is
exact when $V=0$, and gives then in particular for $J=1$ that
\be
\label{eq:ZfreeJ1}
Z=\wt{C}_0^{\;3}\int\rmd^3\bfit{r}_1\;.
\ee
To make this integral finite we should regularize, e.g.\ by introducing
periodic boundary conditions or an external harmonic oscillator potential.
See Section~\ref{sec:Reg} below.

We now take $J$ to be odd, $J=2K+1$, and make the finite Fourier transform
\be
\label{eq:finiteFourier}
\bfit{r}_j
=\sum_{n=-K}^K
\bfit{a}_n\,\rme^{\,\srmi\,{2nj\pi\over 2K+1}}\;.
\ee
It gives that
\be
{S_P\over\hbar}=
\sum_{n=1}^{\infty}C_{n,K}^{\;2}\,|\bfit{a}_n|^2
+{\beta\over J}\sum_{j=1}^J V(\sbfit{r}_j)\;,
\ee
with, for $n=1,2,\ldots$,
\be
C_{n,K}={2(2K+1)\over\hbar}\,\sqrt{m\over\beta}
\,\sin\!\left({n\pi\over 2K+1}\right),\qquad
\wt{C}_{n,K}={C_{n,K}\over\sqrt{\pi}}\;.
\ee
In order to transform the integral over the positions $\bfit{r}_j$ into an
integral over the Fourier components $\bfit{a}_n$, we note that
\be
\sum_j|\rmd\bfit{r}_j|^2
=(2K+1)\left(|\rmd\bfit{a}_0|^2+2\sum_{n=1}^K|\rmd\bfit{a}_n|^2\right),
\ee
and hence,
\be
\prod_j\rmd^3\!\bfit{r}_j
=2^{3K}\left(\sqrt{2K+1}\right)^{3(2K+1)}\,
\rmd^3\!\bfit{a}_0
\prod_{n=1}^K\rmd^6\!\bfit{a}_n\;.
\ee
Using the identity
\be
\sqrt{2K+1}
=2^K\prod_{n=1}^K\sin\!\left({n\pi\over 2K+1}\right),
\ee
we rewrite Equation~(\ref{eq:Zprimitive}) as follows,
\be
\label{eq:ZprimitiveFourier}
Z\approx
\wt{C}_0^{\;3}\int\rmd^3\!\bfit{a}_0
\left(\prod_{n=1}^K
\wt{C}_{n,K}^{\;\;6}\int\rmd^6\!\bfit{a}_n\right)
\exp\!\left(-{S_P\over\hbar}\right).
\ee
The limit $K\to\infty$ gives Equation~(\ref{eq:ZnonfreeinfiniteFourier}). We
regard $\bfit{r}_j$ as a function of the variable $\tau=j\beta\hbar/(2K+1)$,
which becomes continuous in the limit.

\section{Averaging the potential}

The integrand of the path integral,
Equation~(\ref{eq:ZnonfreeinfiniteFourier}), is the negative exponential of
$S/\hbar$, Equation~(\ref{eq:integrandpathint}). One way to interpret this is
that the kinetic part of $S$ defines independent Gaussian probability
distributions for the Fourier coefficients $\bfit{a}_n$, $n>0$, such that the
real and imaginary parts of the $x,y,z$ components of $\bfit{a}_n$ have mean
values zero
and standard deviations
\be
\sigma_n
={1\over\sqrt{2}\,C_n}
={\hbar\over 2n\pi}\,\sqrt{\beta\over 2m}\;.
\ee

The path integral may be computed approximately by the partial
averaging method~\cite{Coalsonetal85,Coalsonetal86}. We integrate
explicitly over the lowest Fourier components, and approximate the
integral over the infinite number of remaining coefficients simply by
averaging the potential. This means that we choose some finite $K$ and
define
\be
\label{eq:defrK}
\bfit{R}(\tau)
=\sum_{n=-K}^K
\bfit{a}_n\,\rme^{\,\srmi\,{2n\pi\tau\over\beta\hbar}}\;.
\ee
The remainder term $\bfit{s}(\tau)=\bfit{r}(\tau)-\bfit{R}(\tau)$ will have
a Gaussian distribution with zero mean and with variances
\be
 \langle(s_x(\tau))^2\rangle
=\langle(s_y(\tau))^2\rangle
=\langle(s_z(\tau))^2\rangle
={\langle(\bfit{s}(\tau))^2\rangle\over 3}
=\sigma^2\;.
\ee
To compute $\sigma$, we compute
\be
\label{eq:tau1tau2correl}
{\langle\bfit{s}(\tau_1)\cdot\bfit{s}(\tau_2)\rangle\over 3}
={\beta\hbar^2\over 2\pi^2m}
\sum_{n=K+1}^{\infty}
{1\over n^2}
\,\cos\!\left({2n\pi(\tau_1-\tau_2)\over\beta\hbar}\right)
={\beta\hbar^2\over 2\pi^2m}\;
f_K\!\left({\tau_1-\tau_2\over\beta\hbar}\right),
\ee
where the function $f_K=f_K(u)$ has period 1 in its argument $u$,
\be
\label{eq:correlfn}
f_K(u)=\sum_{n=K+1}^{\infty}{\cos(2n\pi u)\over n^2}
=\pi^2\left(u-{1\over 2}\right)^2-{\pi^2\over 12}
-\sum_{n=1}^K{\cos(2n\pi u)\over n^2}\;.
\ee
The last formula is valid for $0\leq u\leq 1$. Thus we have
\be
\label{eq:sigmaK}
\sigma^2
={\beta\hbar^2\over 2\pi^2m}\,f_K(0)
={\beta\hbar^2\over 2\pi^2m}\left({\pi^2\over 6}
-\sum_{n=1}^K{1\over n^2}\right)
\approx{\beta\hbar^2\over (2K+1)\pi^2m}\;,
\ee
introducing the approximation
\be
\sum_{n=K+1}^{\infty}{1\over n^2}
\approx\int_{K+{1\over 2}}^{\infty}{\rmd n\over n^2}
={2\over 2K+1}\;,
\ee
which is about 20\% larger than the exact result $\pi^2/6$ in the
worst case $K=0$.

The problem facing us is to compute the integral
\be
I=\prod_{n=K+1}^{\infty}
\left(\wt{C}_n^{\;6}\int\rmd^6\!\bfit{a}_n\;
\exp\!\left(
-C_n^{\;2}\,|\bfit{a}_n|^2\right)\right)\,
\exp\!\left(-{1\over\hbar}\int_0^{\beta\hbar}\rmd\tau\;
V(\bfit{r}(\tau))\right).
\ee
Here $\bfit{r}(\tau)=\bfit{R}(\tau)+\bfit{s}(\tau)$ is given by
Equation~(\ref{eq:infFourierexp}), and $\bfit{R}(\tau)$ by
Equation~(\ref{eq:defrK}). To simplify our notation, we write the integral as
an average. Next, we assume that $\bfit{s}(\tau_1)$ and $\bfit{s}(\tau_2)$ are
uncorrelated for $\tau_1\neq\tau_2$, which is true to a certain approximation,
as shown by Equation~(\ref{eq:tau1tau2correl}). In this approximation we may
compute the integral by the following formal reasoning,
\be
I&\!\!\!=&\!\!\!
\left\langle
\exp\!\left(-{1\over\hbar}\int_0^{\beta\hbar}\rmd\tau\;
V(\bfit{r}(\tau))\right)\right\rangle
=\left\langle
\prod_{\tau=0}^{\beta\hbar}
\exp\!\left(-{\rmd\tau\over\hbar}\,V(\bfit{r}(\tau))\right)
\right\rangle
\nonumber\\
&\!\!\!\approx &\!\!\!
\prod_{\tau=0}^{\beta\hbar}
\left\langle
1-{\rmd\tau\over\hbar}\,V(\bfit{r}(\tau))
\right\rangle
=\exp\!\left(-{1\over\hbar}\int_0^{\beta\hbar}\rmd\tau\;
W(\bfit{R}(\tau))\right),
\ee
where $W$ is an averaged version of the potential $V$,
\be
\label{eq:Wdef}
W(\bfit{R})
=\langle V(\bfit{r})\rangle
=\langle V(\bfit{R}+\bfit{s})\rangle
={1\over(\sqrt{2\pi}\,\sigma)^3}
\int\rmd^3\!\bfit{s}\;\rme^{-{s^2\over 2\sigma^2}}\;
V(\bfit{R}+\bfit{s})\;.
\ee
The standard deviation $\sigma$ is given by Equation~(\ref{eq:sigmaK}). Note
that the effective potential $W$ depends on $K$, since $\sigma$ depends on
$K$.

To summarize, we propose the approximation
\be
Z\approx
\wt{C}_0^{\;3}\int\rmd^3\!\bfit{a}_0
\left(\prod_{n=1}^K \wt{C}_n^{\;6}\int\rmd^6\!\bfit{a}_n\right)
\exp\!\left(-{S_A\over\hbar}\right),
\ee
where
\be
{S_A\over\hbar}=\sum_{n=1}^K C_n^{\;2}\,|\bfit{a}_n|^2
+{1\over\hbar}\int_0^{\beta\hbar}\rmd\tau\;W(\bfit{R}(\tau))\;.
\ee
The most drastic approximation is of course to take $K=0$,
see~\cite{FeynmanHibbs}. Then we get
\be
\sigma=\sqrt{{\beta\hbar^2\over 12m}}\;,
\ee
and
\be
Z\approx
\left({m\over 2\pi\beta\hbar^2}\right)^{3\over 2}
\int\rmd^3\!\bfit{a}_0\;
\exp\!\left(-\beta W(\bfit{a}_0)\right).
\ee

The present version of the partial averaging method is simpler than
the original
one~\cite{Coalsonetal85,Coalsonetal86,Chakravarty93,KoledeRaedt01}, in
that the standard deviation $\sigma$ is taken to be constant. The
method has been used previously for computing the propagator, and not
directly the partition function. Then $\sigma$ has to vary along the
path, since it must vanish at the end points.

\section{Example 1: The harmonic oscillator}

If $V$ is a harmonic oscillator potential,
\be
V(\bfit{r})
={1\over 2}\,m\omega^2\bfit{r}^2\;,
\ee
then the averaged potential $W$ is just $V$ plus a constant,
\be
\label{eq:Wharmosc}
W(\bfit{r})=V(\bfit{r})+{3\over 2}\,m\omega^2\sigma^2\;.
\ee
The addition to the potential contributes a multiplicative factor
in the partition function,
and the resulting approximation is
\be
\label{eq:ZquadrHapprox}
Z\approx\exp\!\left(
-{3(\beta\hbar\omega)^2\over 4\pi^2}
\left({\pi^2\over 6}-\sum_{n=1}^K{1\over n^2}\right)
\right)
{1\over(\beta\hbar\omega)^3}
\prod_{n=1}^K
\left(1+\left({\ns{\beta\hbar\omega}\over\ns{2n\pi}}\right)^2\right)^{-3}\;.
\ee
Another way to obtain the same approximation is to set
\be
1+\left({\beta\hbar\omega\over 2n\pi}\right)^2
\approx
\exp\!\left(\left({\beta\hbar\omega\over 2n\pi}\right)^2\right)
\ee
for $n>K$, in Equation~(\ref{eq:ZquadrH}) with $B=0$, this is valid when we
choose $K$ large enough that
\be
{\beta\hbar\omega\over 2K\pi}<\!\!<1\;.
\ee

\section{Example 2: The Coulomb potential}

Consider now two particles of masses $m_1,m_2$ and charges $q_1,q_2$,
interacting by the Coulomb potential
\be
V=V(\bfit{r}_1,\bfit{r}_2)
={q_1q_2\over 4\pi\epsilon_0\,|\bfit{r}_1-\bfit{r}_2|}\;.
\ee
We write Fourier expansions for the paths of both particles,
\be
\bfit{r}_1(\tau)
=\sum_{n=-\infty}^{\infty}
\bfit{a}_{1n}\,\rme^{\,\srmi\,{2n\pi\tau\over\beta\hbar}}\;,\qquad
\bfit{r}_2(\tau)
=\sum_{n=-\infty}^{\infty}
\bfit{a}_{2n}\,\rme^{\,\srmi\,{2n\pi\tau\over\beta\hbar}}\;.
\ee
The partition function is
\be
\!\!\!\!Z
&\!\!\!=&\!\!\!
\left({\sqrt{m_1m_2}\over 2\pi\beta\hbar^2}\right)^3
\int
\rmd^3\!\bfit{a}_{10}\;
\rmd^3\!\bfit{a}_{20}
\left(\prod_{n=1}^{\infty}
\left({4n^2\pi\sqrt{m_1m_2}\over\beta\hbar^2}\right)^6
\int\rmd^6\!\bfit{a}_{1n}\;\rmd^6\!\bfit{a}_{2n}\right)
\exp\!\left(-{S\over\hbar}\right),
\ee
where
\be
S&\!\!\!=&\!\!\!
\int_0^{\beta\hbar}\rmd\tau\left(
 {1\over 2}\,m_1\,(\dot{\bfit{r}}_1(\tau))^2
+{1\over 2}\,m_2\,(\dot{\bfit{r}}_2(\tau))^2
+V(\bfit{r}_1(\tau),\bfit{r}_2(\tau))
\right)
\nonumber\\
&\!\!\!=&\!\!\!
{4\pi^2\over\beta\hbar}\sum_{n=1}^{\infty}
n^2(m_1\,|\bfit{a}_{1n}|^2+m_2\,|\bfit{a}_{2n}|^2)
+\int_0^{\beta\hbar}\rmd\tau\;V(\bfit{r}(\tau))\;.
\ee
The potential depends only on the relative position
\be
\bfit{r}(\tau)=\bfit{r}_1(\tau)-\bfit{r}_2(\tau)
=\sum_{n=-\infty}^{\infty}
\bfit{a}_n\,\rme^{\,\srmi\,{2n\pi\tau\over\beta\hbar}}\;,
\ee
where $\bfit{a}_n=\bfit{a}_{1n}-\bfit{a}_{2n}$. The real and imaginary parts
of the $x,y,z$ components of the Fourier coefficients $\bfit{a}_n$ have mean
values zero and standard deviations
\be
\sigma_n
=\sqrt{
\langle|\bfit{a}_{1n}|^2\rangle+
\langle|\bfit{a}_{2n}|^2\rangle\over 6}
={\hbar\over 2n\pi}\,\sqrt{\beta\over 2m}\;,
\ee
where $m$ is the reduced mass,
\be
{1\over m}={1\over m_1}+{1\over m_2}\;.
\ee
Like in the one particle case, we define $\bfit{R}(\tau)$ by an equation of
the same form as Equation~(\ref{eq:defrK}). We integrate explicitly over the
Fourier coefficients $\bfit{a}_{1n}$ and $\bfit{a}_{2n}$ up to $n=K$, and we
do the remaining integrations approximately by averaging the potential as in
Equation~(\ref{eq:Wdef}). The averaged Coulomb potential is
\be
\label{eq:WCoulomb}
W(r)
={q_1q_2\over 4\pi\epsilon_0 r}\,
\erf\!\left({r\over\sqrt{2}\,\sigma}\right).
\ee
It equals the Coulomb potential in the limit $r\to\infty$, but is nonsingular
at the origin. The standard deviation $\sigma$ is defined as in
Equation~(\ref{eq:sigmaK}), now with $m$ as the reduced mass. The effect of
the averaging is a multiplication
by the error function, defined as
\be
\erf(x)={2\over\sqrt{\pi}}\int_0^x\rmd u\;\rme^{-u^2}\;.
\ee
%

\section{Numerical computation}

A numerical estimate of the action $S_A$ with the averaged potential $W$ is
the completely discretized action $S_D$ defined by
\be
\label{eq:defSD}
{S_D\over\hbar}=\sum_{n=1}^K C_n^{\;2}\,|\bfit{a}_n|^2
+{\beta\over 2K+1}\sum_{j=0}^{2K}W(\bfit{r}_j)\;,
\ee
where the positions $\bfit{r}_j$ are given by the Fourier
coefficients according to Equation~(\ref{eq:finiteFourier}).

There are at least three arguments in favour of choosing exactly $2K+1$
evaluation points for the action integral of the potential. One is that this
replacement of the integral by a sum is exact for a constant, linear or
quadratic potential. Another argument is that the real Fourier coefficient
$\bfit{a}_0$ and the $K$ complex Fourier coefficients
$\bfit{a}_1,\ldots,\bfit{a}_K$ are just what is needed to fix the $2K+1$
positions $\bfit{r}_j$.

The third argument is less obvious. In fact, our justification of the
averaging procedure defining $W$ suggests that the evaluation points $\tau_j$
should be chosen in such a way that the covariances
$\langle\bfit{s}(\tau_j)\cdot\bfit{s}(\tau_k)\rangle$, given by
Equation~(\ref{eq:tau1tau2correl}), are small. These covariances are
proportional to the function $f_K(u)$ given in Equation~(\ref{eq:correlfn}),
which is symmetric about $u=1/2$, and has $K+1$ zeros between 0 and $1/2$. Of
these zeros, $K$ are close to the values $u_j=j/(2K+1)$, for $j=1,2,\ldots,K$,
as Table~\ref{table1} shows. Thus, with $2K+1$ equally spaced points $\tau_j$
we have that $\langle\bfit{s}(\tau_j)\cdot\bfit{s}(\tau_k)\rangle\approx 0$
for $j\neq k$.

\begin{table}
\begin{center}
\begin{tabular}{|r|ccccc|}
\hline
$K$ & 1 & 2 & 3 & 4 & 5 \\
\hline
0 & 0.211 & & & & \\
1 & 0.265 & 1.068 & & &  \\
2 & 0.273 & 1.110 & 2.032 & & \\
3 & 0.276 & 1.123 & 2.059 & 3.018 & \\
4 & 0.277 & 1.128 & 2.070 & 3.036 & 4.011 \\
5 & 0.278 & 1.131 & 2.075 & 3.045 & 4.024 \\
\hline
10& 0.278 & 1.135 & 2.083 & 3.058 & 4.042 \\
20& 0.279 & 1.136 & 2.086 & 3.061 & 4.047 \\
50& 0.279 & 1.137 & 2.086 & 3.062 & 4.048 \\
\hline
\end{tabular}
\end{center}
\caption{Zeros of the correlation functions $f_K(u)$, with $0<u<1/2$,
multiplied by $2K+1$.}
\label{table1}
\end{table}

This does not necessarily forbid us to use for example twice as many
evaluation points. We may define $\Delta\tau=\beta\hbar/(2K+1)$ and
\be
S_a&\!\!\!=&\!\!\!
\Delta\tau\sum_{j=0}^{2K} W(\bfit{R}(j\Delta\tau))\;,\qquad
S_b=
\Delta\tau\sum_{j=0}^{2K}
W(\bfit{R}((j+\mbox{\scriptsize ${1\over 2}$})\Delta\tau))\;.
\ee
But if we do so, we should perhaps compute
$(\rme^{-{S_a\over\hbar}}+\rme^{-{S_b\over\hbar}})/2$ rather than
$\rme^{-{S_a+S_b\over 2\hbar}}$.

To summarize again, the numerical approximation proposed here is based on
evaluation of the following integral, where $S_D$ is the discrete action
defined in Equation~(\ref{eq:defSD}),
\be
\label{eq:Znumerical}
Z\approx
\wt{C}_0^{\;3}\int\rmd^3\!\bfit{a}_0
\left(\prod_{n=1}^K \wt{C}_n^{\;6}\int\rmd^6\!\bfit{a}_n\right)
\exp\!\left(-{S_D\over\hbar}\right).
\ee

From the partition function we compute the expectation value of the energy as
\be
E=\langle H\rangle
=-\pdd{}{\beta}\,\ln Z(\beta)\;.
\ee
In the above approximate partition function there is $\beta$ dependence in the
coefficients $C_n$ and $\wt{C}_n$, and also in the averaged potential $W$. We
get that
\be
\label{eq:Eapprox}
E\approx {3(2K+1)\over 2\beta}
+{\int\rmd^3\!\bfit{a}_0
\left(\prod_{n=1}^K\int\rmd^6\!\bfit{a}_n\right)
\exp\!\left(-{S_D\over\hbar}\right)
{1\over\hbar}\,\pdd{S_D}{\beta}
\over
\int\rmd^3\!\bfit{a}_0
\left(\prod_{n=1}^K\int\rmd^6\!\bfit{a}_n\right)
\exp\!\left(-{S_D\over\hbar}\right)
}\;,
\ee
and
\be
\label{eq:dSDdbeta}
{1\over\hbar}\,\pdd{S_D}{\beta}
=-{1\over\beta}\sum_{n=1}^K C_n^{\;2}\,|\bfit{a}_n|^2
+{1\over 2K+1}\sum_{j=0}^{2K}
\left(W(\bfit{r}_j)+\beta\,\pdd{}{\beta}\,W(\bfit{r}_j)\right).
\ee
Note that Equation~(\ref{eq:Eapprox}) may be rewritten as
\be
\label{eq:EapproxI}
E\approx {3(2K+1)\over 2\beta}
+{\left(\prod_{j=0}^{2K}\int\rmd^3\!\bfit{r}_j\right)
\exp\!\left(-{S_D\over\hbar}\right)
{1\over\hbar}\,\pdd{S_D}{\beta}
\over
\left(\prod_{j=0}^{2K}\int\rmd^3\!\bfit{r}_j\right)
\exp\!\left(-{S_D\over\hbar}\right)
}\;.
\ee
Here the Fourier coefficients are present only in the kinetic part of the
discrete action $S_D$.

In general we have that
\be
\label{eq:betadWdbeta}
\beta\,\pdd{}{\beta}\,W(\bfit{r})
=\beta\,\pdd{\sigma}{\beta}\,\pdd{}{\sigma}\,W(\bfit{r})
={\sigma\over 2}\,\pdd{}{\sigma}\,W(\bfit{r})\;.
\ee
Hence we get, in the example of the harmonic oscillator potential,
Equation~(\ref{eq:Wharmosc}),
\be
\beta\,\pdd{}{\beta}\,W(\bfit{r})
={3\over 2}\,m\omega^2\sigma^2
\;.
\ee
And in the example of the Coulomb potential, Equation~(\ref{eq:WCoulomb}),
\be
\beta\,\pdd{}{\beta}\,W(\bfit{r})
=-{q_1q_2\over 4\pi\epsilon_0\,(\sqrt{2\pi}\,\sigma)}
\,\exp\!\left({-r^2\over 2\sigma^2}\right).
\ee

In the approximate expression for
$\langle H\rangle=\langle T\rangle+\langle V\rangle$ it is not immediately
obvious which contributions represent kinetic and potential energy,
respectively. In order to identify the different terms, we should define
\be
Z(\beta_1,\beta_2)=\tr\rme^{-\beta_1 T-\beta_2 V}\;,
\ee
and use that, e.g.,
\be
\langle V\rangle=\left. -\pdd{}{\beta_2}\,\ln Z(\beta_1,\beta_2)
\right|_{\beta_1=\beta_2=\beta}\;.
\ee
This formula holds because
\be
\pdd{}{\beta_2}\,\rme^{-\beta_1 T-\beta_2 V}
=-\int_0^1\rmd\lambda\;
\rme^{-\lambda(\beta_1 T+\beta_2 V)}
\,V\rme^{-(1-\lambda)(\beta_1 T+\beta_2 V)}
\;,
\ee
and hence
\be
\pdd{}{\beta_2}\,Z(\beta_1,\beta_2)=
-\tr(V\rme^{-\beta_1 T-\beta_2 V})\;.
\ee
The somewhat surprising conclusion is that
\be
\label{eq:expvalV}
\langle V\rangle\approx
{\left(\prod_{j=0}^{2K}\int\rmd^3\!\bfit{r}_j\right)
\exp\!\left(-{S_D\over\hbar}\right)
{1\over 2K+1}\sum_{j=0}^{2K}W(\bfit{r}_j)
\over
\left(\prod_{j=0}^{2K}\int\rmd^3\!\bfit{r}_j\right)
\exp\!\left(-{S_D\over\hbar}\right)
}\;,
\ee
whereas $\langle T\rangle$ is all the rest of the right hand side of
Equation~(\ref{eq:Eapprox}) or Equation~(\ref{eq:EapproxI}).

For bound states of any number of particles, with the Coulomb
interaction, the virial theorem states that $2\langle T\rangle+\langle
V\rangle=0$. It gives a good check on numerical results for bound
states, if one computes both $\langle T\rangle$ and $\langle
V\rangle$. It may also be used to (potentially) improve the precision
of computed energies, since it implies for example that
\be
E=\langle T\rangle+\langle V\rangle
=-\langle T\rangle={\langle V\rangle\over 2}\;.
\ee
The statistical error with which the two expectation values $\langle T\rangle$
and $\langle V\rangle$ are computed in a Monte Carlo simulation will in
general not be the same, hence one may use whichever value has the smallest
error.

\section{Regularization}
\label{sec:Reg}

For simplicity, we have so far neglected the fact that the partition function
is not mathematically well defined for a system in an infinite volume when,
for example, the potential goes to zero at infinity, like the Coulomb
potential. In our present context, the problem is that the integral over
$\bfit{a}_0$ diverges. In practice, when the integral is computed by some
Monte Carlo method using a random walk algorithm of the Metropolis type, the
divergence means that there is a finite probability of walking away to
infinity, where the potential vanishes. This may be no problem in practice,
because the divergence may be so improbable that it will never happen in the
Monte Carlo simulation. Nevertheless, one may like to introduce some kind of
regularization which makes the partition function well defined.

A convenient regularization method in our case is to add to the Hamiltonian an
extra harmonic oscillator potential
\be
V_0(\bfit{r})={1\over 2}\,m\omega_0^{\;2}\bfit{r}^2\;,
\ee
with a suitably chosen angular frequency $\omega_0$. The Fourier expansion of
Equation~(\ref{eq:infFourierexp}) implies that
\be
\label{eq:pathintregpot}
{1\over\hbar}\int_0^{\beta\hbar}\rmd\tau\;V_0(\bfit{r}(\tau))
={\beta m\omega_0^{\;2}\over 2}\left(
|\bfit{a}_0|^2+2\sum_{n=1}^{\infty}|\bfit{a}_n|^2\right).
\ee
Hence, Equation~(\ref{eq:integrandpathint}) is modified to read
\be
\label{eq:integrandpathintreg}
{S\over\hbar}=
\sum_{n=0}^{\infty}D_n^{\;2}\,|\bfit{a}_n|^2
+{1\over\hbar}\int_0^{\beta\hbar}\rmd\tau\;V(\bfit{r}(\tau))\;,
\ee
where
\be
D_0^{\;2}={\beta m\omega_0^{\;2}\over 2}\;,\qquad
D_n^{\;2}=C_n^{\;2}+\beta m\omega_0^{\;2}
={4(n^2+\nu^2)\pi^2m\over\beta\hbar^2}\;,
\ee
for $n=1,2,\ldots$, and
\be
\nu={\beta\hbar\omega_0\over 2\pi}\;.
\ee
We define also $\wt{D}_n=D_n/\sqrt{\pi}$ for $n=0,1,2,\ldots$. Since
\be
{\wt{C}_0\over\wt{D}_0}\,\prod_{n=1}^{\infty}
{\wt{C}_n^{\;2}\over\wt{D}_n^{\;2}}
={1\over\beta\hbar\omega_0}\prod_{n=1}^{\infty}{n^2\over n^2+\nu^2}
={1\over 2\sinh(\nu\pi)}\;,
\ee
we may rewrite Equation~(\ref{eq:ZnonfreeinfiniteFourier}) as
\be
\label{eq:ZnonfreeinfiniteFourierreg}
Z=Z_0\,
\wt{D}_0^{\;3}\int\rmd^3\!\bfit{a}_0
\prod_{n=1}^{\infty}
\left(\wt{D}_n^{\;6}\int\rmd^6\!\bfit{a}_n\;
\right)
\exp\!\left(-{S\over\hbar}\right),
\ee
where $Z_0$ is the partition function of the three dimensional harmonic
oscillator with angular frequency $\omega_0$,
\be
\label{eq:Z0def}
Z_0=Z_0(\beta)={1\over 8\sinh^3(\nu\pi)}
\;.
\ee
This expression for the partition function $Z=Z(\beta)$ is mathematically well
defined, when the harmonic oscillator potential $V_0$ is included in addition
to the potential $V$, so that Equation~(\ref{eq:integrandpathintreg}) holds.

A natural way to interpret Equation~(\ref{eq:ZnonfreeinfiniteFourierreg}) is
that the Fourier coefficients $\bfit{a}_n$ are Gaussian random variables with
mean zero and standard deviations $\sigma_n=1/(\sqrt{2}\,D_n)$. Our
derivation of how to replace the potential $V$ by an averaged potential $W$,
goes through with little change. The most important change is that the
denominator $n^2$ in Equation~(\ref{eq:tau1tau2correl}) has to be replaced by
$n^2+\nu^2$, and hence the correlation function $f_K$ is replaced by a
function $f_{K,\nu}$ which is still periodic with period 1,
\be
f_{K,\nu}(u)=\sum_{n=K+1}^{\infty}{\cos(2n\pi u)\over n^2+\nu^2}
={\pi\cosh(\nu\pi(2u-1))\over 2\nu\sinh(\nu\pi)}
-{1\over 2\nu^2}
-\sum_{n=1}^K{\cos(2n\pi u)\over n^2+\nu^2}\;,
\ee
the last formula being valid for $0\leq u\leq 1$. The zeros of $f_{K,\nu}$
between 0 and $1/2$ are even closer to the values $u_k=k/(2K+1)$, for
$k=1,2,\ldots,K$, than those of $f_K$, as Table~\ref{table2} shows, for the
arbitrarily chosen value $\nu=10$.

\begin{table}
\begin{center}
\begin{tabular}{|r|ccccc|}
\hline
$K$ & 1 & 2 & 3 & 4 & 5 \\
\hline
0 & 0.055 & & & & \\
1 & 0.113 & 1.003 & & &  \\
2 & 0.151 & 1.011 & 2.002 & & \\
3 & 0.177 & 1.022 & 2.008 & 3.002 & \\
4 & 0.197 & 1.035 & 2.015 & 3.007 & 4.002 \\
5 & 0.211 & 1.048 & 2.021 & 3.012 & 4.006 \\
\hline
10& 0.249 & 1.093 & 2.050 & 3.033 & 4.023 \\
20& 0.269 & 1.122 & 2.073 & 3.051 & 4.038 \\
50& 0.277 & 1.134 & 2.084 & 3.060 & 4.046 \\
\hline
\end{tabular}
\end{center}
\caption{Zeros of the correlation functions $f_{K,\nu}(u)$, with $0<u<1/2$,
multiplied by $2K+1$. The table is for $\nu=10$.}
\label{table2}
\end{table}

The numerical computation now involves the following modified version of
Equation~(\ref{eq:Znumerical}),
\be
\label{eq:Znumericalreg}
Z\approx
Z_0\,\wt{D}_0^{\;3}\int\rmd^3\!\bfit{a}_0
\left(\prod_{n=1}^K \wt{D}_n^{\;6}\int\rmd^6\!\bfit{a}_n\right)
\exp\!\left(-{S_D\over\hbar}\right),
\ee
where we use also a modified definition of the discrete action $S_D$,
\be
{S_D\over\hbar}
&\!\!\!=&\!\!\!
\sum_{n=0}^K D_n^{\;2}\,|\bfit{a}_n|^2
+{\beta\over 2K+1}\sum_{j=0}^{2K}W(\bfit{r}_j)
\nonumber\\
&\!\!\!=&\!\!\!
\sum_{n=1}^K C_n^{\;2}\,|\bfit{a}_n|^2
+{\beta\over 2K+1}\sum_{j=0}^{2K}
\left(V_0(\bfit{r}_j)+W(\bfit{r}_j)\right).
\ee
The standard deviation $\sigma$ to be used in the definition of the averaged
potential $W$, will now be given by the formula
\be
\sigma^2={\beta\hbar^2\over 2\pi^2 m}\,f_{K,\nu}(0)
={\beta\hbar^2\over 2\pi^2 m}\left(
{\pi\coth(\nu\pi)\over 2\nu}
-{1\over 2\nu^2}
-\sum_{n=1}^K{1\over n^2+\nu^2}\right).
\ee

Using Equation~(\ref{eq:Znumericalreg}), we compute the total energy
$\langle E\rangle=\langle T\rangle+\langle V\rangle+\langle V_0\rangle$ as
\be
\label{eq:Eapproxreg}
\langle E\rangle&\!\!\!=&\!\!\!-\pdd{}{\beta}\,\ln Z
\approx
{3(2K+1)\over 2\beta}
+{6\nu^2\over\beta}\,f_{K,\nu}(0)
+{\left(\prod_{j=0}^{2K}\int\rmd^3\!\bfit{r}_j\right)
\exp\!\left(-{S_D\over\hbar}\right)
{1\over\hbar}\,\pdd{S_D}{\beta}
\over
\left(\prod_{j=0}^{2K}\int\rmd^3\!\bfit{r}_j\right)
\exp\!\left(-{S_D\over\hbar}\right)
}\;.
\ee
Compare this to Equation~(\ref{eq:EapproxI}). The modified version of
Equation~(\ref{eq:dSDdbeta}) is the following,
\be
\label{eq:dSDdbetareg}
\!\!\!\!{1\over\hbar}\,\pdd{S_D}{\beta}
=-{1\over\beta}\sum_{n=1}^KC_n^{\;2}\,|\bfit{a}_n|^2
+{1\over 2K+1}\sum_{j=0}^{2K}
\left(V_0(\bfit{r}_j)+W(\bfit{r}_j)
+\beta\,\pdd{}{\beta}\,W(\bfit{r}_j)\right).
\ee
Equation~(\ref{eq:betadWdbeta}) gets modified as follows,
\be
\label{eq:betadWdbetareg}
\beta\,\pdd{}{\beta}\,W(\bfit{r})
=\left(1+\nu\,\pdd{}{\nu}\,\ln f_{K,\nu}(0)\right)
{\sigma\over 2}\,\pdd{}{\sigma}\,W(\bfit{r})\;.
\ee
%
%

In order to calculate separately the expectation values $\langle T\rangle$,
$\langle V\rangle$, and $\langle V_0\rangle$, we should define
\be
Z=Z(\beta_0,\beta_1,\beta_2)=\tr\rme^{-\beta_1 T-\beta_2 V-\beta_0 V_0}\;,
\ee
and keep track of the three parameters $\beta_0,\beta_1,\beta_2$ before
setting them all equal to $\beta$. This gives for $\langle V\rangle$ a formula
exactly like Equation~(\ref{eq:expvalV}). An easier way to compute $\langle
V_0\rangle$ is to note that
\be
\!\!\!\!\langle V_0\rangle
=-{\omega_0\over 2\beta}\,\pdd{}{\omega_0}\,\ln Z
\approx
{3\nu^2\over\beta}\,f_{K,\nu}(0)
+{\left(\prod_{j=0}^{2K}\int\rmd^3\!\bfit{r}_j\right)
\exp\!\left(-{S_D\over\hbar}\right)
{\omega_0\over 2\beta\hbar}\,\pdd{S_D}{\omega_0}
\over
\left(\prod_{j=0}^{2K}\int\rmd^3\!\bfit{r}_j\right)
\exp\!\left(-{S_D\over\hbar}\right)
}\;.
\ee
Here we have that
\be
{\omega_0\over 2\beta\hbar}\,\pdd{S_D}{\omega_0}
={1\over(2K+1)}\sum_{j=0}^{2K}
\left(V_0(\bfit{r}_j)
+{\omega_0\over 2}\,\pdd{}{\omega_0}\,W(\bfit{r}_j)\right),
\ee
with
\be
{\omega_0\over 2}\,\pdd{}{\omega_0}\,W(\bfit{r})
={\omega_0\over 2}\,\pdd{\sigma}{\omega_0}\,
\pdd{}{\sigma}\,W(\bfit{r})
=\left({\nu\over 2}\,\pdd{}{\nu}\,\ln f_{K,\nu}(0)\right)
\left({\sigma\over 2}\,\pdd{}{\sigma}\,W(\bfit{r})\right).
\ee
Once we know $\langle E\rangle$, $\langle V\rangle$ and $\langle V_0\rangle$,
we know also
$\langle T\rangle=\langle E\rangle-\langle V\rangle-\langle V_0\rangle$.
The identification of the various contributions to the total energy
$\langle E\rangle$ is seen to be not entirely trivial.

If $V$ is taken to be the Coulomb potential, the virial theorem gives now that
\be
2\langle T\rangle+\langle V\rangle-2\langle V_0\rangle=0\;.
\ee
This holds for any number of particles. It provides a check on numerical
results, and it may be used to compute the total energy, including the
regulator potential, as
\be
E=\langle T\rangle+\langle V\rangle+\langle V_0\rangle
={\langle V\rangle\over 2}+2\langle V_0\rangle\;.
\ee
The energy including the Coulomb potential but excluding the regulator
potential, is
\be
E_C=\langle T\rangle+\langle V\rangle
={\langle V\rangle\over 2}+\langle V_0\rangle\;.
\ee

\section{Numerical test results}

Table~\ref{table3} presents numerical results for the ground state
energy of the hydrogen atom, for comparison with the exact value of
$-13.598\;$eV. All results are for a temperature of $15\,000\;$K.

The Monte Carlo method was used with a standard Metropolis
algorithm. In each Monte Carlo step, one point to be updated is chosen
randomly among the $2K+1$ points on the discrete path, then a random
step is generated and either accepted or rejected depending on the
change in the discrete action, $\Delta S_D$. If $\Delta S_D\leq 0$,
the step is accepted. If $\Delta S_D>-\hbar\ln u$, with $u$ a uniform
random variable between 0 and 1, the step is rejected.  The
optimization of the Monte Carlo strategy was not considered, but is of
course an important problem. In fact, the naive approach of updating
one point at a time has a disastrously slow convergence when more than
about one hundred time steps are used.

The main computational cost of updating one point is computing the
change in the Fourier components, this takes approximately $2K+1$
floating point operations. If more than one point is updated in each
step, one may choose e.g.\ $2K+1=3^n$ for some power $n$, and then use
the fast Fourier transform with base 3.

By far the highest statistics, $10^{10}$ MC steps, was run for the
entry with 201 time steps.  In this case, statistical uncertainties
are given in the table, and the values found for the ground state
energy are consistent with the exact value, within the uncertainties
of less than one per cent.  A number of time steps of the order of 50
may give sufficient accuracy for many purposes.

It is noteworthy that the statistical error in the direct estimate of
the energy, $\langle T\rangle+\langle V\rangle$, is half the separate
errors in $\langle T\rangle$ and $\langle V\rangle$.  In fact,
$\langle T\rangle+\langle V\rangle$ is seen to be systematically
closer to the exact energy than the estimate $\langle
V_0\rangle+\langle V\rangle/2$ obtained from the virial theorem by
elimination of the kinetic energy.

\begin{table}
\begin{center}
\begin{tabular}{|c|cc|cc|c|cc|}
\hline
No.\ of points &
\multicolumn{2}{c|}{Energies} &
\multicolumn{2}{c|}{Regularization} & Virial &
\multicolumn{2}{c|}{Coulomb energy} \\
$2K+1$ &
$\langle T\rangle$ &
$\langle V\rangle$ &
$\hbar\omega_0$ &
$\langle V_0\rangle$ &
$\Delta$ &
$\langle T\rangle+\langle V\rangle$ &
$\langle V_0\rangle+\langle V\rangle/2$    \\
\hline
\phantom{1}21 & $12.757$ & $-25.486$ & $\phantom{1}0$ & $0.000$ &
$\phantom{-}0.014$ & $-12.729$ & $-12.743$ \\
\phantom{1}41 & $13.063$ & $-26.098$ & $\phantom{1}0$ & $0.000$ &
$\phantom{-}0.014$ & $-13.035$ & $-13.049$ \\
\hline
\phantom{11}1 & $\phantom{0}6.538$ & $-13.193$ &
$\phantom{0}1$ & $0.146$ &
$-0.205$ & $\;\;-6.655$ & $\;\;-6.450$ \\
\phantom{1}11 & $11.676$ & $-23.680$ & $\phantom{0}1$ & $0.064$ &
$-0.228$ & $-12.004$ & $-11.776$ \\
\phantom{1}21 & $12.513$ & $-25.351$ & $\phantom{0}1$ & $0.061$ &
$-0.223$ & $-12.838$ & $-12.615$ \\
\phantom{1}41 & $13.134$ & $-26.747$ & $\phantom{0}1$ & $0.054$ &
$-0.293$ & $-13.613$ & $-13.320$ \\
101 & $13.890$ & $-27.613$ & $\phantom{0}1$ & $0.050$ &
$\phantom{-}0.034$ & $-13.723$ & $-13.757$ \\
201 & $13.489$ & $-26.991$ & $\phantom{0}1$ & $0.056$ &
$-0.063$ & $-13.502$ & $-13.439$ \\
    & $\!\pm 0.141$ & $\;\;\pm 0.181$ & & $\!\!\!\!\!\pm 0.002$ &
$\pm 0.064$ & $\;\;\pm 0.070$ & $\;\;\pm 0.093$ \\
\hline
\phantom{10}1 & $13.417$ & $-20.191$ & $10$ & $3.456$ &
$-0.135$ & $\;\;-6.774$ & $\;\;-6.640$ \\
\phantom{1}11 & $17.728$ & $-28.538$ & $10$ & $3.599$ &
$-0.140$ & $-10.810$ & $-10.670$ \\
\phantom{1}21 & $18.520$ & $-30.363$ & $10$ & $3.461$ &
$-0.122$ & $-11.843$ & $-11.721$ \\
\phantom{1}41 & $18.958$ & $-31.456$ & $10$ & $3.381$ &
$-0.151$ & $-12.498$ & $-12.347$ \\
\hline
\end{tabular}
\end{center}
\caption{
Estimates of the hydrogen ground state
energy.
All values tabulated are in eV.
The quantity
$\Delta=\langle T\rangle-\langle V_0\rangle+\langle V\rangle/2$
should be zero, by the virial theorem.
The last two columns should be compared to the exact value of $-13.598\;$eV.
See comments in the text.}
\label{table3}
\end{table}

\section*{Acknowledgments}

I want to thank the Laboratoire de Physique Th{\'e}orique et
Mod{\`e}les Statistiques at Orsay for their invitation and kind
hospitality. I thank Jean Desbois, Alain Comtet and especially
St{\'e}phane Ouvry for many discussions and useful comments.

\appendix

\section{Example: Quadratic Lagrangian}

In the case of a particle of electric charge $q$ moving in an electromagnetic
vector potential $\bfit{A}(\bfit{r})$, the imaginary time action has also an
imaginary part,
\be
S=\int_0^{\beta\hbar}\rmd\tau\left(
{1\over 2}\,m\,(\dot{\bfit{r}}(\tau))^2
+V(\bfit{r}(\tau))
+\rmi\,q\,\dot{\bfit{r}}(\tau)\cdot
\bfit{A}(\bfit{r}(\tau))\right).
\ee
Note that the contribution from the vector potential is gauge invariant,
because we integrate over a closed path. The partition function can be
computed exactly by the path integral for example when we have an isotropic
harmonic oscillator external potential of angular frequency $\omega$, and a
magnetic field of constant flux density $\bfit{B}$, so that
\be
V(\bfit{r})={1\over 2}\,m\omega^{\;2}\bfit{r}^2\;,\qquad
\bfit{A}(\bfit{r})={\bfit{B}\times\bfit{r}\over 2}\;.
\ee
Then
\be
{S\over\hbar}=
\beta m\omega^2\,|\bfit{a}_0|^2
+\sum_{n=1}^{\infty}
\left(\left(C_n^{\;2}+\beta m\omega^2\right)|\bfit{a}_n|^2
+{2n\pi q\over\hbar}\,
\bfit{B}\cdot(\bfit{a}_n\times\bfit{a}_n^{\ast})
\right),
\ee
and the partition function is, with $B=|\bfit{B}|$,
\be
\label{eq:ZquadrH}
Z=
{1\over(\beta\hbar\omega)^3}
\prod_{n=1}^{\infty}
{\left[
\left(1+\left({\ns{\beta\hbar\omega}\over\ns{2n\pi}}\right)^2\right)
\left(
\left(1+\left({\ns{\beta\hbar\omega}\over\ns{2n\pi}}\right)^2\right)^2
+\left({\ns{\beta\hbar\,|qB|}\over\ns{2n\pi m}}\right)^2
\right)\right]^{-1}}\;.
\ee
The energy spectrum is of course well known. An energy eigenvalue is given by
quantum numbers $j,k,\ell=0,1,2,\ldots$ as
\be
E_{j,k,\ell}
=\left(j+{1\over 2}\right)\hbar\omega_+
+\left(k+{1\over 2}\right)\hbar\omega_-
+\left(\ell+{1\over 2}\right)\hbar\omega\;,
\ee
where
\be
\omega_{\pm}
=\sqrt{\omega^2+\left({|qB|\over 2m}\right)^2}
\pm{|qB|\over 2m}\;.
\ee
Hence,
\be
Z=\sum_{j,k,\ell}\rme^{-\beta E_{j,j,\ell}}
={1\over 8
\sinh\!\left({\ns{\beta\hbar\omega_+}\over\ns{2}}\right)
\sinh\!\left({\ns{\beta\hbar\omega_-}\over\ns{2}}\right)
\sinh\!\left({\ns{\beta\hbar\omega}\over\ns{2}}\right)}\;.
\ee
Equation~(\ref{eq:ZquadrH}) gives a product representation of this function.

\end{document}